\documentstyle[12pt,wspro]{article}
\begin{document}

\title{MICRO--UNIVERSES AND ``STRONG BLACK--HOLES'':\\
A PURELY GEOMETRIC APPROACH TO ELEMENTARY PARTICLES}


\author{E. RECAMI,$^{(*)}$ F. RACITI, W.A. RODRIGUES Jr., and
V.T. ZANCHIN$^{\, (**)}$\\
{\em Dept. of Applied Math., State University at Campinas, S.P., Brazil.}\\
{\em INFN, Sezione di Catania, 57 Corsitalia, 95129--Catania, Italy.}\\
{\em Facolt\`{a} di Ingegneria, Universit\`{a} di Bergamo, Dalmine (BG),
Italy.}
}

\maketitle

\footnotetext{$^{(*)}$ E-mail address in Brazil: \ 47596::Recami
 \ or \ Recami@ccvax.Unicamp.BR}
\footnotetext{$^{\, (**)}$ Present address: Departam. de F\'{\i}sica,
Universidade Federal de Santa Maria, Santa Maria, RG, Brazil.}

\begin{abstract}
We present here a panoramic view of our unified, {\em bi--scale\/} theory of
gravitational and strong interactions \ [which is mathematically analogous to
the last version of N.Rosen's bi--metric theory; and yields physical results
similar to {\em strong gravity\/}'s]. \ This theory, developed during the last
15 years, is purely geometrical in nature, adopting the methods of General
Relativity for the description of hadron structure and strong interactions. \
In particular, hadrons are associated with ``{\em strong\/} black--holes'',
from the external point of view, and with ``micro--universes'', from the
internal point of view. \ Among the results herein presented, let us mention
the derivation: \ (i) of confinement and \ (ii) asymptotic freedom for the
hadron
constituents; \ (iii) of the Yukawa behaviour for the strong potential at the
static limit; \ (iv) of the strong coupling ``constant'', and \ (v) of mesonic
mass spectra.
\end{abstract}

\section{Premise}

Probably each of us, at least when young,  has sometimes imagined that
every small particle of matter could be, at a suitably reduced scale,
a whole cosmos. This idea has very ancient origins. It is already
present, for example, in some works by Democritus of Abdera (about
400~B.C.).  Democritus, simply inverting that analogy, spoke about huge
atoms, as big as our cosmos. And, to be clearer, he added: if one of
those super-atoms (which build up super-cosmoses)
abandoned his  ``giant  universe" to fall down on our world,
our world would be destroyed...

Such kind of considerations are linked to the fantasies about the physical
effects of a  dilation or contraction of all the objects which
surround us, or of the whole ``world''.  Fantasies like these have also
been exploited by several writers: from F.~Rabelais (1565) to
J.~Swift, the narrator of Samuel Gulliver's travels (1727); or to I.~Asimov.
It is probably because of the great diffusion of such ideas that,
when the planetary model of the atom was proposed, it achieved a great
success among people.

Actually, we meet such intuitive ideas in the scientific
arena too.  Apart from the already quoted Democritus, let us
remember the old conception of a {\em hierarchy\/} of
universes ---or rather of cosmoses--- each of them endowed with a particular
{\em scale factor\/} (let us think, for instance, of a series
of russian dolls).  Nowadays, we can really recognize that the
microscopic analysis of matter has revealed grosso modo a series
of ``chinese boxes'':  so that we are entitled to suppose that something
similar may be met also when studying the universe on a large scale, {\em
i.e.},
in the direction of the {\em macro\/} besides of the {\em micro}.
Hierarchical theories were formulated for example by
J.H. Lambert (1761) and, later on, by  V.L. Charlier (1908, 1922)
and F.~Selety (1922--24); followed  more recently  by  O.~Klein,
H.~Alfv\'en and G. de Vaucouleurs, up to the works of A.~Salam and
co-workers, K.P. Sinha and C.~Sivaram, M.A. Markov, E.~Recami and
colleagues, D.D. Ivanenko and collaborators, M.~Sachs, J.E. Charon,
H.~Treder, P.~Roman, R.L. Oldershaw, and others.\cite{Ref-1}

\section{Introduction}

In this paper we confine ourselves to examine the possibility  of
considering elementary particles as micro universes:\cite{Ref-2}
that is to say, the possibility that they be similar ---in a
sense to be specified--- to our cosmos.  More precisely, we shall refer
ourselves to the thread followed by  P.~Caldirola, P.~Castorina,
A.~Italiano, G.D. Maccarrone, M.~Pavsic, and ourselves.\cite{Ref-3}

 Let us recall that Riemann, as well as  Clifford and later
Einstein,\cite{Ref-4} believed that the fundamental particles of matter
were the
perceptible evidence of a strong local space curvature.  A theory
which stresses the role of space (or, rather, space-time)
curvature already does exist for our whole cosmos: General Relativity,
based on Einstein gravitational field equations; which are
probably the most important equations of
classical physical theories, together with Maxwell's electromagnetic
field equations.
Whilst much effort has already been  made to generalize
Maxwell equations, passing for example from the electromagnetic field
to Yang--Mills fields (so that almost all modern gauge theories are
modelled on Maxwell equations), on the contrary Einstein equations
have never been applied to domains different from the gravitational
one.  Even if they, as any differential equations,
{\em do not\/} contain any inbuilt fundamental length: so that they
can be used a priori to describe cosmoses of any size.

Our first purpose is now to explore how far it is possible to apply
successfully the  methods of general relativity
(GR), besides to the world of gravitational interactions, also to
the domain of the so--called nuclear, or  {\em strong\/},
interactions:\cite{Ref-5} namely, to the world of the elementary
particles called hadrons. A second purpose is linked to the fact that
the standard theory (QCD) of strong interactions has not yet fully
explained why the hadron constituents (quarks) seem to be permanently
{\em confined\/} in the interior of those particles; in the sense
that nobody has seen up to now an isolated ``free'' quark, outside
a hadron. So that, to explain that confinement, it has been
necessary to invoke phenomenological models, such us the so--called
``bag'' models, in their MIT and SLAC versions for instance. The
``confinement'' could be explained, on the contrary, in a natural way and
on the basis of a well--grounded theory
like GR, if we associated with each hadron (proton, neutron, pion,...) a
particular ``cosmological model''.

\section{The Model by Micro-Universes}

Let us now try to justify the idea of considering the strong interacting
particles (that is to say, hadrons) as micro-universes.
We meet a first motivation if we think of the so--called ``large
number coincidences'',  already known since several decades and
stressed by  H.~Weyl, A.I. Eddington, O.~Klein, P.~Jordan, P.A.M.
Dirac, and by others.

The most famous among those empirical  observations is that the
ratio $R/r$ between the radius  $R\simeq 10^{26}{\rm m}$ of our cosmos
(gravitational universe) and the typical radius $r\simeq 10^{-15}{\rm
m}$ of elementary particles is grosso modo equal to the ratio
$S/s$ between the strength $S$ of the nuclear (``strong'') field
and the strength $s$ of the gravitational field (we will give
later a definition of $S$, $s$):
$$
\rho \equiv {R \over r} \simeq {S \over s} . \eqno(1)
$$

This does immediately suggest the existence of a {\em similarity},
in a geometrico--physical sense, between cosmos and hadrons. As a
consequence of such similarity, the ``theory of models''
yields ---by exploiting simple dimensional considerations--- that,
if we contract our cosmos of the quantity
$$
\rho = R/r \approx 10^{41}
$$
(that is to say, if we transform it in a hadronic micro-cosmos
{\em similar\/} to the previous one), the field strength would increase
in the same ratio: so
to get  the gravitational field transformed into the strong one.

If we observe, in addition, that the typical duration of a  decay is
inversely proportional to the strength of the interaction itself, we
are also able to explain why the mean-life of our gravitational cosmos
($\Delta t \simeq  10^{18}$ s: duration ---for example--- of a complete
expansion/contraction cycle, if we accept the theory of the cyclic
{\em big bang\/}) is a multiple, with the same ratio,  of
the typical mean-life ($\Delta \tau \simeq 10^{-23}{\rm s}$) of the
``strong micro-universes", or hadrons:
$$
\Delta t \simeq \rho \Delta \tau . \eqno(2)
$$

It is also interesting that, from the self-consistency of
these deductions implies ---as we shall show later--- that the mass
$M$ of our cosmos should be equal to $\rho ^{2} \simeq (10^{41})^2$
times the typical mass $m$ of a hadron: \ a fact that   seems to
agree with reality, and constitutes a further ``numerical
coincidence'', the so--called Eddington relation. Another numerical
coincidence is shown and explained in ref.\cite{Ref-6}

By making use of Mandelbrot's language\cite{Ref-7} and of his
general equation for  self-similar structures, what precedes can be
mathematically translated into the claim that
cosmos and  hadrons are systems, with scales
$N$ and $N-1$,  respectively, whose ``fractal dimension'' is
$D = 2$, where $D$ is the auto-similarity exponent that characterizes
the hierarchy. As a consequence of all that, we shall assume that
cosmos and hadrons (both of them regarded of course as finite objects)
be {\em similar\/} systems: \ that is, that
they be governed by similar laws, differing only for
a ``global'' scale transformation which  transforms $R$ into $r$ and
gravitational field into strong field. [To fix our ideas, we may
{\em temporarily\/} adopt the na\"\i ve model of a ``newtonian ball''
in three--dimensional space for both cosmos and hadrons.  Later on, we shall
adopt more sensible models, for example Friedmann's].
Let us add, incidentally, that we should be ready {\em a priori\/} to accept
the
existence of other cosmoses besides ours: \ let us recall that
man in every epoch has successively called  ``{\em universe\/}'' his
valley, the whole Earth, the solar system, the Milky Way and today
(but with the same simple--mindedness) our cosmos, as we know it on the
basis of our  observational and theoretical
instruments\ldots\cite{Ref-8}

Thus, we arrive at a {\em second\/} motivation for our theoretical
approach: That physical laws should be covariant (= form invariant) under
{\em global\/} dilations or contractions of space-time.
We can easily realize this if we notice that:
(i) when we dilate (or contract) our measure units of space and time,
physical laws, of course, should {\em not\/} change their form; \  (ii) a
dilation of the measure units is totally equivalent to a contraction
(leaving now ``meter" and ``second" unaltered) of the
observed world.

Actually, Maxwell equations of electromagnetism ---the most important
equations of classical physics, together with Einstein equations,
as we already said--- are by themselves covariant
also under conformal transformations and, in particular, under
dilations. In the case when electric charges are present, such a
covariance holds provided that charges themselves are suitably
``scaled''.

Analogously, also Einstein {\em gravitational\/} equations are
covariant\cite{Ref-9} under dilations: provided that, again, when in
the presence of matter and of a cosmological term $\Lambda$, they too
are scaled according to correct dimensional considerations. \
The importance of this fact had been well realized by Einstein himself,
who two weeks before his death wrote, in connection with his last
unified theory: \
$<<$From the form of the field  [{\em gravitational\/} +
{\em electromagnetic\/}]  equations it follows immediately that:
if $g_{i k}(x)$
is a solution of the field equations, then also $g_{i k}(x/\alpha)$,
where $\alpha$ is a positive constant, is a solution
(``similar solutions'').  Let us suppose, for example, that
$g_{i k}$ represents a finite  crystal embedded in a flat space.
It is then possible that a second `universe' exists with another
crystal, identical with the first one, but
dilated  $\alpha$ times with respect to the former.
As far as we confine ourselves to consider a universe containing
only one crystal, there are no difficulties: we just realize that the
size of such a crystal (standard of length) is not determined by the
field equations\ldots$>>$.
These lines are taken from Einstein's preface to the Italian book
{\em Cinquant'anni di Relativit\`a}.\cite{Ref-10} They have been
written in Princeton on April 4th, 1955, and stress the fact,
already mentioned by us, that  differential equations ---as all the
fundamental equations of physics--- do not contain any inbuilt
``fundamental length''.  In fact,  Einstein equations can  describe
the internal dynamics of our cosmos, as well as of much bigger
super-cosmoses, or of much smaller  micro-cosmoses (suitably ``scaled'').
\begin{figure}
\vspace*{12. cm}
\begin{center}
\parbox{13. cm}{\small {\bf Fig.~1} --  {\em ``Coloured" quarks and their
strong charge} -- This scheme represents the complex
plane\cite{Ref-3,Ref-12,Ref-13} of the {\em sign\/}  $s$ of
the quark strong--charges $g_{\rm j}$ in a hadron. These strong charges
can have $three$ signs, instead of two as in the case of the ordinary
electric charge $e$.  They can be represented, for instance, by \ $s_{1} =
(i-\sqrt{3})/2$; $s_{2} = (i+\sqrt{3})/2$; $s_{3} = -i$, \ which
correspond to the arrows separated by $120^{\rm o}$ angles. \
The corresponding anti-quarks will be endowed with strong charges
carrying the complex conjugate signs \ ${\overline
{s}}_{1}$, ${\overline{s}}_{2}$, ${\overline{s}}_{3}$. \ The three quarks
are represented by the ``yellow" (Y), ``red" (R) and ``blue" (B)
circles; the three anti-quarks by the ``violet" (V), ``green" (G) and
``orange" (O) circles.  The latter are complementary to the former
corresponding colors. \ Since in real particles the inter--quark
forces
are saturated, hadrons are white. The white colour can be obtained either
with three--quark structures, by the combinations YRB or VGO (as it happens in
baryons and antibaryons, respectively), or with two--quark structures, by the
combinations YV or RG or BO [which are actually quark--antiquark combinations],
as it happens in mesons and their antiparticles. See also note\cite{Ref-11}.
}
\end{center}
\end{figure}
\section{A Hierarchy of  ``Universes''}

As a first step for better exploiting the symmetries of the
fundamental equations of classical physics, let us therefore fix our
attention on the space-time {\em dilations\/}
$$
x'_{\mu} = \rho x_{\mu}  \eqno(3)
$$
with $x_{\mu} \equiv (t;x,y,z)$ and $\mu =0,1,2,3$, and explicitly
require physical laws to be covariant with respect to them: under the
hypothesis, however, that only  {\em discrete\/} values of $\rho$ are
realized in nature. As before, we are moreover supposing that
$\rho$ is constant as the space or time position varies (global, besides
{\em discrete\/}, dilations).

Let us recall that natural objects interact essentially through four
 (at least) fundamental forces, or interactions:
the gravitational, the ``weak'', the electromagnetic  and the
``strong'' ones; here
listed according to their (growing) strength.
 It is possible to express such strengths by pure numbers, so to be
allowed to compare them each other.  For instance, if one chooses to
define each strength as the dimensionless {\em square\/} of a
``vertex coupling
constant'', the electromagnetic strength results to be measured
by the (dimensionless) coefficient \ $Ke^{2}/\hbar
\equiv \alpha \simeq 1/137$, where $e$  is the electron charge,
 $\hbar$ the reduced Planck constant, $c$ is the light speed in
vacuum and $K$ is the electromagnetic interaction universal constant
(in the International System of units,
$K=(4\pi \varepsilon_{0})^{-1}$, with $\varepsilon_{0}$ =  vacuum
dielectric constant). \ Here we are interested in particular
in the gravitational and strong interaction strengths:
$$
s \equiv Gm^{2}/\hbar c ; \qquad  S \equiv Ng^{2}/\hbar c ,
$$
where  $G$ and $N$ are the gravitational and
strong universal constants, respectively;  quantities $m$ e $g$
representing  the gravitational charge (=mass) and the strong
charge\cite{Ref-11,Ref-12} (cf.~Fig.~1), respectively, of one and the same
hadron:
for example of a nucleon  $\cal N$ or of a pion $\pi$. More precisely, we
shall often adopt in the following the convention of calling  $m$ and $g$
``{\em gravitational mass\/}''  and ``{\em strong mass\/}'',
respectively.

Let us consider, therefore,  two  identical particles endowed with both
gravitational ($m$) and strong ($g$) mass, {\em i.e.}, two
identical hadrons, and the ratio between the strengths  $S$ and $s$
of the corresponding strong and gravitational interactions.  We find \
 $S/s \equiv Ng^{2}/Gm^{2} \simeq 10^{40 \div 41}$, \ so that one
verifies that \
 $\rho \equiv R/r \simeq S/s$. \ For
example for $m=m_{\pi}$ one gets $Gm^{2}/\hbar c \simeq 1.3 \times
10^{-40}$, while the pp$\pi$ or $\pi \pi \rho$ (or quark-quark-gluon:
see below) coupling constant squares are \ $Ng^{2}/\hbar c \simeq 14$ or 3
(or 0.2), respectively.

Already at this point, we can make some simple remarks.  First of all, let us
notice that, {\em if we put\/} conventionally $m \equiv g$, then the
strong universal constant $N$ becomes
$$
N \simeq \rho G \approx hc/m_{\pi}^{2}. \eqno(4)
$$

On the contrary, if {\em we choose\/} units such that
$[N] = [G]$ and moreover $N = G  = 1$, we obtain
$g = m \sqrt{\rho}$  and, more precisely (with  $n=2$ or $n=3$),
$$
g_{\rm o} = g/n \simeq \sqrt{\hbar c/G} \equiv {\rm  Planck} \;\;
{\rm mass},
$$
which tells us that ---in suitable units---
the so--called ``Planck mass'' is nothing but the {\em magnitude}
of the rest strong--mass [= strong charge] of a typical hadron, or  rather
of quarks.\cite{Ref-11}

{}From this point of view, we should {\em not\/} expect
the ``micro black--holes" (with masses of
the order of the Planck mass), predicted by various Authors, to exist; \
in fact, we {\em already} know of
the existence of quarks, whose {\em strong charges\/} are of
the order of the Planck mass (in suitable units). \  Moreover, the
fact ---well known in standard theories---  that gravitational
interactions become as strong as the ``strong" ones for masses of the
order of the Planck mass does simply mean in our opinion that the {\em strong
gravity\/} field generated by quarks inside hadrons (strong micro-universes)
is nothing but the strong nuclear field.

\section{``Strong Gravity''}

A consequence of what stated above is that inside
 a hadron ({\em i.e.},  when  we want to describe strong interactions
among hadron constituents) it must be possible to adopt the same
Einstein equations which are used for the description of
gravitational interactions inside our cosmos; with the only warning of
scaling
them down, that is, of suitably {\em scaling\/}, together with space distances
and time durations, also the gravitational constant  $G$ (or the
masses) and the cosmological constant $\Lambda$.

Let us now recall that Einstein's equations for gravity do essentially state
the equality of two tensorial quantities: the first describing the geometry
(curvature) of space-time, and the second ---that we shall call ``matter
tensor'', $G T^{\mu \nu}$--- describing the distribution of matter:
$$
R_{\mu \nu} - {1 \over 2} g_{\mu \nu} R^{\rho}_{\rho} -
\Lambda g_{\mu \nu} = -k G T_{\mu \nu} ; \qquad
[k \equiv {8 \pi \over c^{4}}].  \eqno(5)
$$
As well-known, \  $G \simeq 6.7 \times 10^{-11}
{\rm m}^{3}/({\rm kg}\times {\rm s}^{2})$, while $\Lambda \approx
10^{-52} {\rm m}^{-2}$.

Inside a hadron, therefore, equations of the same form will hold, except
that instead of $G$ it will appear (as we already know) quantity
$N \approx hc/m_{\pi} ^{2}$ and instead of $\Lambda$ it will appear
the  ``strong cosmological constant'' (or ``hadronic constant'')
$\lambda$:
$$
N \equiv \rho_{1} G ; \qquad  \lambda \equiv \rho^{2} \Lambda;
\qquad \rho_{1} \approx \rho, \eqno(6)
$$
so that $\lambda \simeq 10^{30} {\rm m}^{-2} = (1 \; {\rm fm})^{-2}$,
or ${\lambda}^{-1} \approx 0.1$  barn.

For brevity's sake, we shall call $S_{\mu \nu} \equiv
N T_{\mu\nu}$ the ``strong matter tensor''.

What precedes can be directly applied, with  a
satisfactory  degree of approximation, to the case ---for example--- of
the pion: {\em i.e.}, to the case of the cosmos/pion similarity. Almost as
if our cosmos were a
super--pion, with a  super--quark (or ``metagalaxy'', adopting
Ivanenko's terminology) of matter and one of anti-matter. Let us
recall however that, as we already warned in Section~{\bf 3}, the
parameter $\rho$ {\em can\/} vary according to the
particular cosmos and hadron considered.  Analogously
$\Lambda$, and therefore  $\lambda$, can vary too: \ with the further
circumstance that a priori
also their {\em sign\/} can change, when varying the object (cosmos or
hadron) taken into examination.

As far as $\rho_{1}$ is concerned, an even more important remark has to be
made.
Let us notice that the gravitational coupling constant $Gm^{2}/\hbar c$
(experimentally measured in the case of the interaction of two
``tiny components'' of our particular cosmos) should be compared with
the analogous
constant for the interaction of two tiny {\em components\/}
(partons? partinos?) of the corresponding hadron, or rather of a
particular constituent quark  of its.  That constant is unknown to us.
We know
however, for the simplest hadrons, the quark-quark-gluon coupling
constant:  $Ng^{2}/\hbar c \simeq 0.2$.  As a consequence, the best
value for $\rho_{1}$ we can predict  ---up to now---  for those hadrons is
$\rho_{1} \simeq 10^{38} \div 10^{39}$ [and, in fact, $10^{38}$ is the
value which has provided the results most close  to the
experimental data]: \ a value that however will vary, let us repeat it,
with the particular cosmos and the particular hadron chose for the comparison.

The already mentioned ``large numbers'' empirical relations, which link
the micro- with the macro-cosmos, have been obtained by us as a
{\em by-product\/} of our scaled--down equations for the interior  of
hadrons, and of the ordinary Einstein equations.  Notice, once
more, that our ``numerology'' connects the gravitational
interactions with the strong ones, and {\em not} with the electromagnetic ones
(as Dirac, instead, suggested).  It is worthwhile noticing that  strong
 interactions, as the gravitational ---but differently from the
 electromagnetic ones,--- are highly non-linear and then associable
to {\em non}-abelian gauge theories.  One of the purposes of
our theoretical approach consists, incidentally,  in proposing an
{\em ante litteram\/}  geometrical interpretation  of those theories.

Before going on, let us specify that the present  {\em geometrization\/}
of the strong field is justified by the circumstance that the ``Equivalence
principle"
(which recognizes the identity, inside our cosmos, of
inertial and gravitational mass)  can be extended
to the hadronic universe in the following way.
The usual Equivalence principle can be understood, according to Mach,
thinking of the inertia  $m_{\rm I}$ of a given body as
due to its interaction with all the other masses of the
universe: an interaction which in {\em our\/} cosmos is essentially
gravitational; so that  $m_{\rm I}$ coincides with the gravitational
mass: $m_{\rm I} \equiv m_{\rm G}$. Inside a ``hadronic cosmos'',
however, the predominant interaction among its constituents is the
strong one; so that the inertia $m_{\rm I}$ of a constituent
will coincide with its strong charge $g$ (and not with $m_{\rm G}$).
We  shall see that our generalization of the Equivalence principle
will be useful for geometrizing the strong field not only inside a hadron,
but also in its neighborhood.

Both for the cosmos and for hadrons, we shall adopt Friedmann--type
models; taking advantage of the fact that they
are compatible with the Mach Principle, and are  embeddable
in 5~dimensions.

\section{In the Interior of a Hadron}

Let us see some {\em consequences\/} of our Einstein--like
equations, re-written for the strong field and therefore valid inside
a hadron:
$$
R_{\mu \nu} - {1 \over 2} g_{\mu \nu} R^{\rho}_{\rho} -
\lambda g_{\mu \nu}  =  -k S_{\mu \nu} ; \qquad
[S_{\mu \nu} \equiv N T_{\mu \nu}]. \eqno(7)
$$

In the case of a spherical constituent, that is to say of a spherically
symmetric distribution $g'$ of ``strong mass'', and in the usual
Schwarzschild-deSitter $r$,$t$ coordinates, the known geodesic motion
equations for a small test--particle  (let us call it a {\em parton},
with strong mass $g"$) tell us that it will feel
a ``force'' easy to calculate,\cite{Ref-3,Ref-13} which for low
speeds [{\em  static limit\/}: $v<<c$] reduces to the (radial) force:
$$
F = -{1 \over 2} c^{2} g"
(1 - {{2Ng}' \over {c^{2} r}} + {1 \over 3}\lambda r^{2})({{2Ng'}\over
{c^{2}r^{2}}} + {2 \over 3} \lambda r). \eqno(8)
$$

Notice that, with proper care, also in the present case one can introduce a
language in terms of ``force'' and ``potential'';  for example in Eq.~8 we
defined \ $F \equiv g"{\rm d}^{2}r/{\rm d}t^{2}$. \ In Fig.~2
the form is depicted of two typical potentials yielded by the present
theory [cf. Eq.~8'].

\begin{figure}[htb]
\vspace*{10.3 cm}
\begin{center}
\parbox{13. cm}{\small {\bf Fig.~2} --   In this figure the shape is shown of
two typical  inter--quark
potentials $V_{\rm eff}$ yielded by the present theoretical approach: cf.
Eq.~8.
We show also the theoretical energy--levels {\em calculated\/} for the
${1-}^{3}s_{1}$, ${2-}^{3}s_{1}$ e ${3-}^{3}s_{1}$ states of ``Bottomonium"
and ``Charmonium", respectively [by adopting for the {\em bottom\/} and
{\em charm\/} quark the masses $m$(b)=5.25 and $m$(c)=1.68 GeV/$c^{2}$]. \
The comparison with experience is satisfactory:\cite{Ref-17}  see
Section~{\bf 5}.
}
\end{center}
\end{figure}

At  ``intermediate distances'' ---i.e., at the newtonian
limit--- this force simply reduces to $F \simeq -{1 \over 2}
c^{2}g"(2Ng'/c^{2}r^{2} + 2 \lambda r/3)$, that is, to
the sum of a newtonian term and of an elastic term  {\em \`a la\/}
Hooke.  Let us notice that, in such a limit, the last expression is
valid even when the test particle $g"$ does {\em not\/} posses a
small strong mass, but is ---for example--- a second quark.
Otherwise, our expressions for $F$ are valid only
{\em approximately\/} when also  $g"$ is a quark; nevertheless,
they can explain some important features of the hadron constituent
behaviour, both for small and for large values of  $r$.

At very large distances, when  $r$ is of the same order of (or is greater
than) the considered hadron radius \
[$r \geq  \; \sim \! \! 10^{-13}$ cm $\equiv 1 \; {\rm fm}$], \ whenever
we confine ourselves to the simplest hadrons  (and thus choose
$\Lambda \simeq 10^{30}$ m$^{-2}$; $N \simeq 10^{38 \div 39} G$),
we end with an {\em attractive\/} radial force which is proportional
to  $r$:
$$
F \approx -g"c^{2} \lambda r/3. \eqno(9)
$$

In other words, one naturally obtains a confining force
(and a confining potential $V \div r^{2}$) able a priori to
explain the so--called {\em confinement\/} of the hadron
constituents (in particular, of quarks).  Because of this force, the
motion of $g"$ can be regarded in a  first approximation as a
harmonic motion; so that our {\em theory\/} can include the various
and interesting results already found by different Authors for the hadronic
properties  ---for instance, hadron mass spectra---
just by {\em postulating\/} such a motion.

Up to now we supposed $\lambda$ to be positive.  But it is
worthwhile noticing that confinement is obtained also for negative
values of $\lambda$.  In fact, with less drastic approximations,
for $r \geq \; \sim \! \! 1 \; {\rm fm}$ one gets:
$$
F \approx -{1 \over 3}g"c^{2} \lambda (r + \lambda r^{3}/3 -
Ng'/c^{2}), \eqno(9')
$$
where, for $r$ large enough, the ${\lambda}^{2}$ term is dominating.
Let us warn however that,  when considering ``not simple'' hadrons
(so that $\lambda$, and moreover $N$, may change their values), other
terms can become important, like the newtonian one, $-N g'^{2} /
r^{2}$, or even the {\em constant\/} term $+N \lambda g'^{2} /3$ which
corresponds to a linear potential.  Let us observe, finally, how this
last equation predict that, for inter--quark distances of the order of
1~fm, two quarks have to attract each other with a force of {\em some
tons\/}: a quite huge force, especially when recalling that it should act
between  two extremely tiny
particles (the {\em constituents} of mesons and
baryons), whose magnitute would increase with the distance.

Let us pass to consider, now, {\em not} too big distances, always at the
static limit.  It is then important
to add to the radial potential the usual
``kinetic energy term'' (or centripetal potential), $(J/g")^{2}/2r^{2}$,
in order to account for the orbital angular momentum of $g"$ with
respect to $g'$.  The effective potential\cite{Ref-13}
between the two constituents $g'$, $g"$  gets thus the following form
$$
V_{\rm eff} = {1 \over 2}g"c^{2} [2({{Ng'} \over {c^{2}}})^{2}
{1 \over r^{2}} - {{2Ng'} \over {c^{2}}}{1 \over r} - {{2\lambda Ng'}
\over {3c^{2}}}r + {\lambda \over 3}r^{2}
+ {1 \over 2}{({\lambda \over 3})^{2}}r^{4}] + {{(J/g")^{2}}
\over {2r^{2}}}, \eqno(8')
$$
which, in the region where GR reduces essentially to the newtonian theory,
simplifies into:
$$
V_{\rm eff} \approx  -Ng'g"/r + (J/g")^{2}/2r^{2}.
$$

In such a case the test particle  $g"$ can set itself (performing a
circular motion, for example: and in Section~{\bf 7}
we shall give more details) at a distance $r_{\rm e}$ from
the source--constituent at which  $V$ is minimum; {\em i.e.}, at
the distance \ $r_{\rm e} =J^{2}/Ng'g"^{2}$. \ At this
distance the ``effective force'' vanishes.
Thus we meet, at short distances, the phenomenon known as
{\em asymptotic freedom\/}: For not large distances (when the force
terms proportional to $r$ and to $r^{3}$ become negligible), the
hadron constituents behave as if they were (almost) free. \ If we
now extrapolated, somewhat arbitrarily, the expression for  $r_{\rm e}$
to the case of two quarks [for example, $|g'| = |g"| = g_{\rm o}
\simeq {1 \over 3} m_{\rm p}$], we would obtain the preliminary
estimate $r_{\rm e} \approx {1 \over 100} \, {\rm fm}$.
Vice-versa, by supposing ---for instance in the case of baryons, with
$g \equiv m \simeq m_{\rm p}$ and $N \simeq 10^{40} G$--- that the
equilibrium radius $r_{\rm e}$ be of the order of a
hundredth of a fermi, one would  get the Regge--like relation
$J/\hbar \simeq m^{2}$ (where  $m$ is measured in GeV/$c^{2}$).

Let us perform these calculations again, however, by using the {\em complete}
expression of
$V_{\rm eff}$.  First of all, let us observe  that it is  possible to
evaluate the radius at which the potential reaches its minimum also
in the case $J = 0$. \
By extrapolation to the case of the simplest quarks
[for which $Ng^{2}/\hbar c \simeq 0.2$], one finds
always at least one solution, $r_{\rm e} \approx  0.25 \; {\rm fm}$,
for $\lambda$ positive and of the order of $10^{30}$ m$^{-2}$. \  Passing
to the case $J = \hbar$ (which corresponds classically
to a speed $v \simeq c$ for the moving quark), we obtain under the
same hypothesis the value
$$
r_{\rm e} \simeq 0.9 \ {\rm fm}.
$$
Actually, for positive  $\lambda$ it exists  the
above solution {\em only\/}. \
For negative values of $\lambda$,  however, the situation is more
complex; let us summarize it in the case of the $N$ and
$| \lambda |$ values adopted by us.  One meets ---again--- at least one
solution, which for   $J = 0$ takes the simple analytic form
${r_{\rm e}}^{3} = 3Ng'/c^{2}| \lambda |$.

More precisely, for $\lambda = -10^{30}$m$^{-2}$ one finds the values
0.7 and 1.7 fm, in correspondence to $J=0$ and $J=1$; \ values that
however become 0.3 and 0.6 fm, respectively, for $\lambda =
-10^{29} {\rm m}^{-2}$. \ In the  $J = 0$ case, at last,
two {\em further\/} solutions are met, the smaller one
[for $\lambda = -10^{30}$ m$^{-2}$] being once more $r_{\rm e}
\simeq 0.25$ fm.

By recalling that  {\em mesons\/} are made up of two quarks
(q, \=q), our approach suggests for mesons in their ground
state ---when $J = 0$, at least--- the model of two quarks {\em
oscillating\/} around an equilibrium position.
It is rather interesting to notice that for small oscillations
(harmonic  motions in space)  the dynamical group would then be SU(3).
It is interesting to notice, too, that the value
$m_{\rm o} = h \nu /c^{2}$, corresponding
to the frequency $\nu = 10^{23}$ Hz,  yields  the pion mass:
$m_{\rm o} \simeq m_{\pi}$.

Analogous results have to hold, obviously, for our cosmos
(or, rather, for the cosmoses which are ``dual'' to the hadrons
considered).

\section{The Strong Coupling Constant}

Here we want  to add just that, in the case of a spherically
symmetric, static metric (and in the coordinates in which it is
diagonal), the Lorentz factor is proportional to $\sqrt{g_{oo}}$, so
that the {\em  strong coupling constant\/}  $\alpha_{\rm S} \equiv S$
in our theory\cite{Ref-14} assumes the form:\cite{Ref-15}
$$
\alpha_{\rm S} (r) \simeq {N \over \hbar c} {{{{g'}_{\rm o}}^{2}}
\over {1 - 2N{g'}_{\rm o}/c^{2}r + \lambda r^{2}/3}}, \eqno(10)
$$
since the strong mass $g"$ depends on the speed:
$$
g" = {{{g"}_{\rm o}} \over \sqrt{g_{oo}}} = {{{g"}_{\rm o}} \over
\sqrt{1 - 2N{{g'}_{\rm o}}/r + \lambda r^{2} / 3}}, \eqno(11)
$$
so as the ordinary relativistic mass does.  The behaviour of our
``constant'' $\alpha_{\rm S} (r)$ is analogous to  that one of the
perturbative coupling constant of the ``standard theory''  (QCD):
that is to say, $\alpha_{\rm S} (r)$ decreases as  the distance $r$
decreases, and increases as it increases, once more
justifying both confinement and ``asymptotic freedom''. \
Let us recall that, when\cite{Ref-15} ${g"}_{\rm o} = {g'}_{\rm o}$,
the definition of $\alpha_{\rm S}$ is \ $\alpha_{\rm S} \equiv S = N
{g'}^{2} / \hbar c$.

Since the Schwarzschild--like coordinates ($t;r,\theta,\varphi$)
do not correspond, as is well known, to any real observer,
it is interesting from the {\em physical\/} point of view to pass to the
local coordinates ($T;R, \theta, \varphi$) associated with
observers who are  {\em at rest\/} ``with respect to the metric'' at
each point  ($r, \theta, \varphi$) of space: \ ${\rm d}T \equiv
\sqrt{g_{tt}} {\rm d}t$; ${\rm d}R \equiv \sqrt{-g_{rr}}{\rm d}r$,
where $g_{tt} \equiv g_{oo}$ \ and \ $g_{rr} \equiv g_{11}$. \ These
{\em ``local'' observers\/} measure a speed  $U \equiv {\rm d}R/{\rm d}T$
(and  strong masses)  such that
$\sqrt{g_{tt}} = \sqrt{1 - U^{2}}$, so that Eq.~11 assumes
the transparent form
$$
g" = {{{g"}_{\rm o}} \over {\sqrt{1 - U^{2}}}}. \eqno(11')
$$

Once  calculated (thanks to the geodesic equation) the speed
$U$ as a function of  $r$, it is easy to find  again, for example, that
for negative $\lambda$ the minimum value of  $U^{2}$ corresponds to
$r = [3N{{g'}_{\rm o}}/|\lambda|]^{1/3}$. \ While for positive $\lambda$
we get a similar expression,  i.e., \ $r_{\rm o} \equiv
[6N{{g'}_{\rm o}}/\lambda]^{1/3}$, \ which furnishes a limiting
({\em confining\/}) value of $r$, which cannot be reached by
any of the constituents.

Let us finally consider the case of a geodesic circular motion, as
described by the  ``physical'' observers, {\em i.e.}, by our local
observers (even if we  find it convenient to express everything as a
function of the old Schwarzshild-deSitter coordinates). If $a$ is
the angular momentum per unity of  strong rest-mass, in the case of
a test--quark in motion around the source--quark, we meet the
interesting relation $g" = {{g'}_{\rm o}} \sqrt{1 + a^{2}/r^{2}}$,
which allows us to write the strong coupling constant in the
particularly simple form\cite{Ref-14}
$$
\alpha_{\rm S} \simeq {{N} \over {\hbar c}}{{g'}_{\rm o}}(1 + {{a^{2}}
\over {r^{2}}}). \eqno(10')
$$

We can now observe, for instance, that ---if $\lambda < 0$--- the
specific angular momentum $a$ vanishes in correspondence to the customary
geodesic \ $r \equiv r_{\rm qq} = [3N{{g'}_{\rm o}}/|\lambda|]^{1/3}$; \
in this case the test--quark can remain {\em at rest}, at a distance
$r_{\rm qq}$ from the source--quark. With the ``typical'' values
$\rho = 10^{41}; \ \rho_{1} = 10^{38}$, and  ${{g'}_{\rm o}} =
m_{\rm p}/3 \simeq 313$ MeV/$c^{2}$, we obtain $r_{\rm qq} \simeq
0.8 \; {\rm fm}$.

\section{Outside a Hadron. Strong Interactions among Hadrons}

{}From the ``external'' point of view, when describing
the interactions among hadrons (as they appear to us in {\em our\/}
space), we are in need of {\em new\/} field equations able to account for
both the gravitational and strong field which surround a hadron.
We need actually a {\em bi-scale\/} theory [Papapetrou], in order to study
for example the motion in the vicinity of a hadron of a test--particle
possessing both
gravitational and strong mass.

What precedes suggests ---as a first step--- to represent the strong
field around a source--hadron  by means of a tensorial field,
$s_{\mu \nu}$, so as it is tensorial (in GR) the
gravitational field
$e_{\mu \nu}$. Within our theory,\cite{Ref-3,Ref-2,Ref-1} Einstein
gravitational equations have been actually {\em modified\/} by
introducing, in the neighborhood of a hadron, a strong
deformation  $s_{\mu \nu}$  of the metric,
acting only on objects having a strong charge
({\em i.e.}, an intrinsic  ``scale factor'' $f \simeq 10^{-41}$) and not
on objects possessing only a gravitational charge ({\em i.e.}, an
intrinsic  scale factor $f \simeq 1$).  Outside a hadron, and for a
``test--particle'' endowed with both the charges, the {\em new\/} field
equations are:
$$
R_{\mu \nu} + \lambda s_{\mu \nu} =
 -{{8 \pi} \over {c^{4}}} \ [S_{\mu \nu} - {1 \over 2}g_{\mu \nu}
S^{\rho}_{\rho}].  \eqno (12)
$$

They reduce to the usual Einstein equations
far from the source--hadron, because they  {\em imply\/} that the
strong field exists only in the very neighborhood of the hadron: namely that
(in suitable coordinates) \  $s_{\mu \nu} \rightarrow \eta_{\mu \nu}$
for $r >> 1 \; {\rm fm}$.

\noindent{\bf Linear approximation:} -- For distances from the
source--hadron $r \geq \; \sim \! \! 1 \; {\rm fm}$,
when our new field equations can be linearized, the total metric
$g_{\mu \nu}$ can be written as the {\em sum\/} of the two
metrics $s_{\mu \nu}$ \ and \ $e_{\mu \nu}$; or, more precisely
(in suitable coordinates):
$$
2g_{\mu \nu} = e_{\mu \nu} + s_{\mu \nu} \simeq \eta_{\mu \nu} +
s_{\mu \nu}.
$$
Quantity $s_{\mu \nu}$ can then be written as \
$s_{\mu \nu} \equiv \eta_{\mu \nu} + 2 h_{\mu \nu}$, \ with
$|h_{\mu \nu}| << 1$; so that
$g_{\mu \nu} \simeq \eta_{\mu \nu} + h_{\mu \nu}$ (where,
let us repeat, $h_{\mu \nu} \rightarrow 0$ per $r>>1 \; {\rm fm}$).
For the sake of simplicity, we are in addition confining ourselves to
the case of positive $\lambda$  [on the contrary,
if  $\lambda < 0$, we should\cite{Ref-13}  put \ $s_{\mu \nu} \equiv
\eta_{\mu \nu} - 2 h_{\mu \nu}$].

One of the most interesting results is that, at the static limit
(when only $s_{oo} \not = 0$ and the strong field becomes a scalar
field), we get that \
$V \equiv h_{oo} \equiv {1 \over 2} (s_{oo} - 1) = g_{oo} - 1$ \ is
exactly the {\em Yukawa potential\/}:
$$
V = -g {{{\rm exp}[- \sqrt{2 |\lambda|} r]} \over r} \simeq -
{g \over r} {{\rm exp}[{{-m_{\pi} r c} \over \hbar}]}, \eqno(13)
$$
with the correct coefficient ---within a factor  2--- also in the
exponential.\cite{Ref-3,Ref-2,Ref-1}

\noindent{\bf Intense field approximation:} -- Let us consider the
source--quark as an {\em axially\/} symmetric  distribution of strong charge
$g$: the study of the metrics in its neighborhood will lead us to
consider a Kerr-Newman-deSitter (KNdS)--like problem and to look for
solutions of the type ``{\em strong\/} KNdS black holes''.
We find that ---from the ``external'' point of view--- hadrons can be
associated with the above mentioned ``{\em strong black-holes\/}''
(SBH), which result to have radii  $r_{\rm S} \approx 1 \; {\rm fm}$.

For $r \rightarrow r_{\rm S}$, that is, when the field is very
intense, we can perform  the approximation just ``opposite'' to the linear
one, by assuming  $g_{\mu \nu}\simeq s_{\mu \nu}$.  We obtain, then,
equations which are essentially identical with the ``internal''
ones [which is good for the matching of the hadron interior and exterior!];
a consequence being that what we are going to say can be valid
also for {\em quarks}, and not only for hadrons. \
Before going on, let us observe that $\lambda$ can {\em a priori\/}
take a certain sign outside a hadron, and the opposite sign inside it.
In the following we shall confine ourselves to the case $\lambda < 0$
for simplicity's sake.

In general for negative  $\lambda$ one meets\cite{Ref-14} three
``strong horizons'', {\em i.e.}, three values of $r_{\rm S}$, that we
shall call $r_{1}$, $r_{2}$, $r_{3}$. \ If we are interested in hadrons
which are {\em stable\/}  with respect to the strong interactions,
we have to look for those solutions for which the SBH
Temperature\cite{Ref-16} [= strong field strength at its surface]
almost vanishes.  It is worth noticing that the condition of a
vanishing field at the SBH surface implies the coincidence of two, or
more, strong horizons;\cite{Ref-3,Ref-14,Ref-16} and that such
coincidences  imply in their turn some ``Regge--like'' relations  among
$m$, $\lambda$, $N$, $q$ and $J$, \ if $m$, $q$, $J$ are ---now---
mass, charge and {\em intrinsic angular momentum\/} of the
considered hadron, respectively. More precisely, if we choose a priori
the values of $q$, $J$, $\lambda$ and $N$, then our theory yields
{\em mass\/} and  {\em radius\/} of the corresponding stable hadron.
Our theoretical approach is, therefore, a rare example of a formalism
which can yield ---at least a priori--- the {\em masses\/} of the stable
particles (and of the quarks themselves).

\section{Mass Spectra}

We arrived at the point of checking whether and how our approach can yield
the values of the hadron
masses and radii: in particular for hadrons stable
with respect to strong interactions; one can guess a priori that
such values will possess the correct order of magnitude.  Several
calculations have been performed by us, in particular for the meson
mass spectra;\cite{Ref-13,Ref-14} although they ---because of our laziness
with respect to numerical elaborations--- are still waiting for
being reorganized.

Here we quickly outline just some of the  results.
At first, let us consider the case of the simultaneous coincidence
of all the {\em three\/} horizons ($r_{1} = r_{2} = r_{3} \equiv
r_{\rm h}$).  We get a system of equations that ---for example--- rules
out the possibility that  intrinsic angular momentum  (spin) $J$ and
electric charge  $q$ be simultaneously zero
[{\em practically\/} ruling out particles with $J = 0$]; \ it also implies
the interesting relation \ ${\lambda}^{-1} \simeq 2{r_{\rm h}}^{2}$; \
and finally it admits (real and positive) solutions only for {\em low} values
of $J$, the upper limit of the spin depending on the chosen parameters.

The values we obtained for the (small) radii and for the masses
suggest that the ``{\em triple coincidences\/}'' represent the case of
{\em quarks}. The basic formulae for the explicit calculations are
the following.\cite{Ref-14} First of all, let us put $N = \rho_{1} G$,
so that $g \equiv m$.  Let us then define, as usual, \ $Q^{2} \equiv
Nq^{2}/Kc^{4}$; \ $a \equiv J/mc$; \ $M \equiv Nm/c^{2}$, and moreover \
$\delta \equiv 1 + \lambda a^{2}/3$. Then, the radii of the stable
particles (quarks, in this case) are given by the {\em simple\/}
equation \ $r = 3M/2\delta$; \ but the masses are given by the
solution of a  system of two Regge--like relations: \ $9M^{2} =
-2\delta^{3}/\lambda$;  \ $9M^{2} = 8\delta (a^{2} +
Q^{2})$.

The cases of ``{\em double coincidence}'', that is, of the coincidence of
two (out of three) horizons only, seem to be able to describe stable
baryons and mesons. The fundamental formulae become, however, more
complex.\cite{Ref-14} Let us define \ $\eta \equiv a^{2} + Q^{2}$; \
$\sigma \equiv \delta^{2} + 4\lambda \eta$; \
$Z \equiv 3\delta^{2} - 4\lambda \delta \eta + 18 \lambda M^{2}$. \
The stable hadron's radii are then given by the relation
$r \equiv 3M\sigma/Z$; while the masses are given by the non simple
equation $9M^{2}\sigma (\delta \sigma - Z) + 2\eta Z^{2} = 0$,
which relates $M$ with $a$, $Q$ and $\lambda$.  Of course,  some
simplifications are met in particular cases.
For example, when $\lambda = 0$,  we  get the Regge--like relation:
$$
M^{2} = a^{2} + Q^{2}, \eqno(14)
$$
which ---when $q$ is negligible--- becomes $M^{2} = cJ/G$,
that is [with $c=G=1$]:
$$
m^{2} = J . \eqno(14')
$$
On the contrary, when $J = 0$, and $q$ is still negligible, we obtain
[always with $c=G=1$]:
$$
9 m^{2} = -{\lambda}^{-1}. \eqno(15)
$$
Also in the cases of  ``triple coincidence'' simple expressions are found,
when $|\lambda a^2| << 1$. \ Under such a condition, one meets
the simple system of two equations:
$$
9M^{2} \simeq 8(a^{2} + Q^{2}); \qquad 9m^{2} \simeq -2 {\lambda}^{-1},
\eqno(16)
$$
where the second relation is written with $c=G=1$.

All the  ``geometric'' evaluations of this Section~{\bf 9} are
referred  ---as we have seen--- only to  {\em stable} hadrons
 ({\em i.e.}, to hadrons corresponding to SBHs with
 ``temperature" $T \simeq 0$), because we do not know of general rules
associating a temperature $T$ with the many {\em resonances} experimentally
discovered (which will correspond\cite{Ref-1,Ref-2,Ref-3} to temperatures
{\em of the order } of $10^{12}$K, if they have to
 ``evaporate" in times  of the order  of $10^{-23}$s). \
Calculations apt at comparing our theoretical approach with experimental
mass spectra (for mesons, for example) have been till now performed, therefore,
by making recourse to the trick of inserting  our inter--quark potential
$V_{\rm eff}$, found in  Section~{\bf 6}, into a Schroedinger equation. \
Also such (many) calculations
 \ ---kindly performed by our colleagues Prof. J.A. Roversi e Dr. L.A.
Brasca--Annes of the ``Gleb Wataghin"  Physics Institute of the
State University at Campinas (S.P., Brazil)--- \ have not yet been reordered!
\ Here let us specify, nevertheless,  that potential  (8') has been inserted
into the Schroedinger equation in spherical (polar) coordinates, which has
been solved by a finite difference method.\cite{Ref-13}

In the case of  ``Charmonium" and of  ``Bottomonium", for example,
the results obtained (by adopting\cite{Ref-17} for the quark masses the values
$m$(charm) = 1.69 GeV/$c^{2}$; \ $m$(bottom) = 5.25 GeV/$c^{2}$) are
the following (Fig.~2). \ For the states ${1-}^{3}s_{1}$, ${2-}^{3}s_{1}$
and ${3-}^{3}s_{1}$  of Charmonium, we obtained the energy levels
  3.24, 3.68 and 4.13 GeV, respectively.
\ Instead, for the corresponding quantum states of Bottomonium,
we  obtained the energy levels   9.48, 9.86 and 10.14 GeV, respectively.
 \ The radii for the two fundamental states resulted to be \
 $r$(c)=0.42$\;$fm, \ and \ $r$(b)=0.35$\;$fm,
\ with $r$(c)$>r$(b) \ [as  expected  from
 ``asymptotic freedom"]. \ Moreover, the  values of the parameters obtained
by our  computer
fit  are actually those expected: \ $\rho = 10^{41}$ \ and \ $\rho_{1} =
10^{38}$
 \ (just the ``standard" ones) for  Charmonium; \ and $\rho = 0.5 \times
10^{41}$ \ and \ $\rho_{1} = 0.5 \times 10^{38}$ for  Bottomonium.

The correspondence between experimental and theoretical
 results\cite{Ref-17} is satisfactory, especially when recalling the
approximations adopted (in particular, the one of treating the second quark
$g"$
as a test--particle).

\vspace*{1. cm}

\section{Acknowledgements}

This work was partially supported by CNPq, and by INFN,
CNR and MURST. \ The authors are grateful, for many useful discussions
or for the kind collaboration offered over the years, to
P.Ammiraju, L.A. Brasca--Annes, A. Bugini, P. Castorina, A. Garuccio,
L. Lo Monaco, A. Italiano, G.D. Maccarrone, J.E. Maiorino, R. Mignani,
M. Pav\v{s}i\v{c},  J.A. Roversi, M. Sambataro,
F. Selleri, A. Taroni, V. Tonini,  and  ---in particular---
to P. Picchi, E.C. de Oliveira, S. Sambataro, Q. de Souza and M.T.
Vasconselos. \ At last,  one of us (FR)
gratefully acknowledges a grant from the ``Fondazione Bonino--Pulejo'',
Messina (Italy).

\vfill\eject


\section{References}

\begin{thebibliography}{17}
\leftmargin 2.5em

\bibitem{Ref-1} See for example A.~Salam and D.~Strathdee: {\em Phys. Rev.}
D{\bf 16} (1977) 2668; D{\bf
18} (1978) 4596;  A.~Salam: in {\em Proceed. 19th Int. Conf. High-Energy
Physics
(Tokio,1978)}, p.937; {\em Ann. N.Y. Acad. Sci.} {\bf 294} (1977) 12;
C.~Sivaram
and K.P. Sinha: {\em Phys. Reports\/} {\bf 51} (1979) 111;  M.A. Markov:
{\em Zh. Eksp. Teor.
Fiz.} {\bf 51} (1966) 878; E.~Recami and P.~Castorina: {\em Lett. Nuovo Cim.}
 {\bf 15}
(1976) 347; R.~Mignani: {\em ibidem} {\bf 16} (1976) 6;
  P.~Caldirola, M.~Pavsic and
E.~Recami: {\em Nuovo Cimento} B{\bf 48} (1978) 205; {\em Phys. Lett.}
A{\bf 66} (1978) 9;
P.~Caldirola and E.~Recami: {\em Lett. Nuovo Cim.} {\bf 24} (1979) 565;
D.D. Ivanenko:
in {\em Astrofisica e Cosmologia, Gravitazione, Quanti e Relativit\`a --
Centenario di Einstein}, edited by M.~Pantaleo and F. de Finis (Giunti-Barbera;
Florence, 1978), p.131;  R.L. Oldershaw: {\em Int. J. General Systems} {\bf 12}
(1986) 137. See also N.~Rosen: {Found. Phys.} {\bf 10} (1980) 673.

\bibitem{Ref-2} See for example  E.~Recami: in {\em Old and New Questions
in Physics,
Cosmology,...},  by A. van der Merwe (Plenum; New York, 1983); {\em Found.
Phys.} {\bf 13} (1983) 341.  Cf. also P.~Ammiraju, E.~Recami and W.A.
Rodrigues:
{\em Nuovo Cimento} A{\bf 78} (l983) 172.

\bibitem{Ref-3} For an extended summary of that theory, see for example
E.~Recami: {\em Prog. Part.
Nucl. Phys.} {\bf 8} (1982) 401, and refs. therein; \  E.~Recami,
J.M. Mart\'\i nez and V. Tonin--Zanchin: {\em Prog. Part. Nucl. Phys.} {\bf 17}
(1986) 143; \ and E.~Recami and V. Tonin--Zanchin: {\em Il Nuovo Saggiatore}
{\bf 8} (1992; issue no.2) 13. \
See also E.~Recami and V. Tonin--Zanchin: {\em Phys. Lett.} B{\bf 177}
(1986) 304; B{\bf 181} (1986) E416.

\bibitem{Ref-4} See for example A.~Einstein: ``Do gravitational fields play
an essential
role in the structure of elementary particles?", {\em Sitzungsber. d. Preuss.
Akad. d. Wiss.}, 1919 (in German).

\bibitem{Ref-5} See also M.~Sachs: {\em Found. Phys.} {\bf 11} (1981) 329;
 \ and {\em General Relativity and Matter} (Reidel; Dordrecht, 1982).

\bibitem{Ref-6} A.~Italiano and E.~Recami: {\em Lett. Nuovo Cim.} {\bf 40}
(1984) 140.

\bibitem{Ref-7} See for example B.B. Mandelbrot: {\em The Fractal Geometry
of Nature} (W.H.Freeman; San Francisco, 1983).

\bibitem{Ref-8} This clarifies that our
geometrico--physical similarity holds between two classes of objects
of different scale (hadrons and cosmoses), in the sense that the factor
 $\rho$ will vary according to the particular
cosmos and hadron considered.  That will be
important for the practical applications.
At last, let us recall that in Mandelbrot's philosophy,
analogous objects do exist at every hierarchical level, so that
 we can  conceive a particular type of cosmos for each particular type
of hadron, and vice-versa. As a consequence, we should expect
$\rho$  to change a little in each case
 (for example, according to the hadron type considered).

\bibitem{Ref-9} Le us notice that we do not refer here
to the usual  ``general covariance"  of the Einstein's equations
 (that are supposed to hold in our cosmos), but to their covariance with
 respect to transformations ({\em dilations}) which bring  --for example--
from our cosmos to the hadronic micro-cosmos.

\bibitem{Ref-10} M.~Pantaleo (editor): {\em Cinquant'anni di Relativit\`a}
(Giunti; Firenze, 1955).

\bibitem{Ref-11} Let us recall that the hadron constituents
 (2 for mesons and 3 for baryons) have named
 {\em quarks} by M.~Gell-Mann.  This Anglo-Saxon word, which usually means mush
 or also curd, is usually ennobled by literary quotations
(for example, Gell-Mann  was inspired  ---as it is well known---
 by a verse of J.~Joyce's {\em  Finnegans wake}, 1939).  Let us here quote that
Goethe had properly used such a word in his {\em Faust}, verse 292,
where Mephistopheles referring to mankind  exclaims:
$<<$In Jeden Quark begr\"abt er seine Nase"$>>$!\hfill \break
By considering  quarks to be the real carriers of the strong charge
(cf. Fig.~1), we can call  ``colour" the {\em sign}  $s_{j}$ of such
strong charge;$^{12}$  namely, we can regard hadrons as
 endowed with a
zero total strong charge,  each quark possessing the strong
 charge $g_{j} = s_{j}|g_{j}|$  with
$\Sigma s_{j} = 0$.  Therefore, when passing from ordinary
gravity to  ``strong gravity", we shall replace  $m$ by
$g = ng_{\rm o}$, quantity
$g_{\rm o}$  being the average  {\em magnitude} of the constituent quarks
 {\em  rest}--strong-charge,
and $n$  their number.$^{12}$

\bibitem{Ref-12} Cf. for example E. Recami: in {\em Annuario '79, Enciclopedia
EST-Mondadori}, ed. by  E. Macorini (Mondadori; Milan, 1979), p.59.

\bibitem{Ref-13} A.~Italiano {\em et al.}: {\em Hadronic J.}  {\bf 7}
(1984) 1321; \
P.~Ammiraju {\em et al.}: {\em Hadronic J.} {\bf 14} (1991) 441; \
V. Tonin--Zanchin: M.Sc. Thesis (UNICAMP; Campinas, S.P., 1987); \ E.~Recami
and
V.T. Zanchin: ({\em in preparation}).

\bibitem{Ref-14} V. Tonin--Zanchin, E.~Recami, J.A. Roversi and L.A.
Brasca--Annes: ``About
some Regge--like relations for (stable) black holes", report IC/91/219
(ICTP; Trieste, 1991), to appear in {\em Comm. Theor. Phys.}; \ E.~Recami
and V. Tonin--Zanchin: ``The strong coupling constant: Its theoretical
derivation
from a geometric approach to hadron structure", {\em Comm. Theor. Phys.} {\bf
1}
(1991) 111.

\bibitem{Ref-15} Actually, if we considered
a (light)  test--particle $g"$ in the field of a
 ``heavy" constituent $g'$ (a quark for instance), we would rather obtain
 only a square root at the denominator; namely $\alpha_{\rm S} \simeq
({N {g'}_{\rm o} {g"}_{\rm o}} / {\hbar c}) [\sqrt{1 -
2 N {g"}_{\rm o} /c^{2} r +
\lambda r^{2} / 3}]^{-1}$. \  When we pass to consider  two
heavy constituents (two quarks) endowed with the same  {\em rest} strong-mass
${g"}_{\rm o} = {g'}_{\rm o}$, we ought to tackle
the two body problem in GR; however, in an approximate way, and looking at an
 {\em average} situation, one can   propose a formula like  Eq.~13,
where  $r$ is the distance from the common
 ``centre of  mass".

\bibitem{Ref-16} See for example J.D. Bekenstein: {\em Phys. Rev.} D{\bf 9}
(1974) 3292;  \ S.W. Hawking:
{\em Comm. Math. Phys.} {\bf 43} (1975) 199.

\bibitem{Ref-17} C.~Quigg: report 85/126--T (Fermilab, Sept. 1985).

\end{thebibliography}
\end{document}